\documentclass[prd,aps,nofootinbib,superscriptaddress]{revtex4}
\usepackage{epsfig}
\usepackage{amsmath, slashed}
\usepackage{xcolor}
\usepackage{appendix}
\usepackage{dsfont}

\usepackage[normalem]{ulem}

\usepackage{bm}
\usepackage{amsfonts}
\allowdisplaybreaks
\begin{document}

\title{Novel relations for twist-3 tensor-polarized fragmentation functions in spin-1 hadrons}

\author{Qin-Tao Song}
\email[]{songqintao@zzu.edu.cn}
\affiliation{School of Physics and Microelectronics, Zhengzhou University, Zhengzhou, Henan 450001, China}

\date{\today}

\begin{abstract}
{ There are three types of fragmentation functions (FFs) which are used to describe the twist-3 cross sections of the hard semi-inclusive processes under QCD collinear factorization, and they are called intrinsic, kinematical, and dynamical FFs.
 In this work, we investigate the theoretical relations among these FFs for a tensor-polarized spin-1 hadron.
Three Lorentz-invariance relations  are derived by using the identities between the nonlocal quark-quark and quark-gluon-quark operators, which guarantee the frame independence of the twist-3 spin observables. The QCD equation of motion relations are also presented for the tensor-polarized FFs.
In addition, we show that the intrinsic and kinematical  twist-3 FFs can be decomposed into the contributions of twist-2 FFs and twist-3 three-parton FFs, and the latter are also called dynamical FFs. If one neglects the dynamical FFs, we can obtain relations which are analogous to the Wandzura-Wilczek relation. Then, the intrinsic and kinematical twist-3 FFs are expressed in terms of the leading-twist ones. Since the FFs of a spin-1 hadron can be measured at various experimental facilities in the near future, these theoretical relations will play an important role in the analysis of the collinear tensor-polarized FFs.}

\end{abstract}

\maketitle

\date{}

\section{Introduction}
\label{introduction}
Parton distribution functions (PDFs) are key physical quantities in hadron spin physics, since they are used to
solve the proton spin puzzle and to understand the inner structure of hadrons. For a spin-1/2 hadron, the  theoretical relations of PDFs and fragmentation functions (FFs)
have been well studied. Starting with the Wandzura-Wilczek (WW) relation, it is known that
if one neglects the three-parton PDFs, the twist-3 PDF $g_2$ can be expressed in terms of the leading-twist one $g_1$ which has been well measured
~\cite{Wandzura:1977qf}. The violation of the WW relation comes from the three-parton PDFs, and it was shown that such  violation  can be as large as 15\%-40\% of the size of $g_2$~\cite{Accardi:2009au}.
There also exist the so-called Lorentz-invariance relations (LIRs) for the PDFs in a spin-1/2 hadron, which were investigated in Refs.~\cite{Mulders:1995dh, Belitsky:1997ay, Tangerman:1994bb, Kundu:2001pk,Goeke:2003az, Metz:2008ib, Accardi:2009au,Kanazawa:2015ajw}.
In addition to PDFs, LIRs were also derived  for the quark FFs~\cite{Kanazawa:2015ajw}.
Recently, the authors of Ref.~\cite{Koike:2019zxc} performed a systematic study  on  the gluon PDFs and FFs, where the  intrinsic and kinematical twist-3 gluon distributions are written in terms of the twist-2 distributions and the
twist-3 dynamical distributions, and the latter are actually three-parton distributions; moreover, the complete LIRs are also listed for the gluon part.
On the one hand, these interesting relations can be used as constraints for the analysis of twist-3 distributions. On the other hand, they are also crucial to describe
the spin observables, for example, the LIRs  can be used  to guarantee the  frame independence of the twist-3 cross sections, such as the single-spin
asymmetries (SSAs) in the hadron production of lepton-nucleon collisions and the hadron production of  hadronic collisions ($pp\rightarrow \Lambda^{\uparrow} X$)~\cite{Kanazawa:2015ajw,Koike:2017fxr,Koike:2021awj}.

For a  spin-1 hadron, there are unpolarized, vector-polarized and tensor-polarized distributions. The former two also exist for a spin-1/2 hadron, while  the tensor-polarized distributions are the new ones. Among the tensor-polarized PDFs, $b_1(x)$ [or $f_{1LL}(x)$]~\cite{Hoodbhoy:1988am, Frankfurt:1983qs} and the gluon transversity  $\Delta_Tg(x)$~\cite{Jaffe:1989xy,Nzar:1992ax} are the most interesting ones. The sum rule of $\int dx b_1(x)=0$ was derived for an isoscalar object such as the deuteron, and the breaking of this sum rule is related to the contribution of a tensor-polarized component of the sea quarks and antiquarks~\cite{Close:1990zw}. In 2005, the HERMES collaboration performed
 the first measurement of $b_1(x)$ for deuteron with slightly large uncertainties~\cite{HERMES:2005pon}, and it indicates that $b_1(x)$ is much larger than the theoretical prediction~\cite{Cosyn:2017fbo}.
Since  the theoretical estimate of $b_1(x)$ was given  by considering deuteron as a weakly bound state of proton and neutron, the large $b_1(x)$ could indicate exotic components of deuteron such  as a six-quark state and a hidden color state~\cite{Miller:2013hla}.
As for the gluon transversity $\Delta_Tg(x)$, it is related to the helicity flipped amplitude, so it only exists in a hadron  with spin more than or equal to 1 due to the angular momentum conservation. In this case, one can infer that there are nonnucleonic components in the deuteron by the nonzero $\Delta_Tg(x)$, which means that it is interesting to investigate the gluon transversity by experiment; for example, it can be extracted from the cross sections of deep-inelastic scattering (DIS)~\cite{Jaffe:1989xy, Ma:2013yba} and Drell-Yan process~\cite{Kumano:2019igu,Kumano:2020gfk} with a tensor-polarized deuteron target. In the near future, $b_1(x)$ and $\Delta_Tg(x)$ will be measured
at the Thomas Jefferson National Accelerator Facility (JLab)~\cite{jlab-b1,jlab-b2}, Fermilab (Fermi National Accelerator Laboratory) \cite{Keller:2020wan,Keller:2022abm,Clement:2023eun}, and Nuclotron-based Ion Collider fAcility (NICA)\cite{Arbuzov:2020cqg}.  There are also interesting theoretical relations for the tensor-polarized PDFs; in Ref.~\cite{Kumano:2021fem} the twist-3 PDF $f_{LT}(x)$ was decomposed into the contributions of a twist-2 PDF $b_1(x)$ [$f_{1LL}(x)$] and the three-parton PDFs.
Moreover, the WW-type relation was obtained by dropping the latter. The QCD equation of motion (e.o.m.) relations and LIR were derived in Ref.~\cite{Kumano:2021xau} for tensor-polarized PDFs.
Recently, the gluon transversity generalized parton distribution  was also investigated for a spin-1 hadron~\cite{Cosyn:2018rdm}, which becomes the gluon transversity  $\Delta_Tg(x)$ in the forward limit. In addition to the collinear PDFs, one can find the tensor-polarized transverse-momentum dependent (TMD) PDFs   up to twist 4 for a spin-1 hadron in Refs.~\cite{Bacchetta:2000jk,kumano:2021,Ninomiya:2017ggn,Boer:2016xqr}.

The spin-1 hadrons are produced in the hard semi-inclusive processes, such as $\rho$, $\phi$, $K^{\ast}$ and a deuteron.
In order to describe those processes, the tensor-polarized FFs are needed. The quark collinear FFs are defined in Ref.~\cite{Ji:1993vw} up to twist 4 for a spin-1 hadron, and
the tensor-polarized TMD FFs can be also found in Refs.~\cite{Bacchetta:2000jk,Chen:2016moq}.
In the future, the tensor-polarized FFs can be measured at BESIII and Belle II. Actually, such measurement is now in progress, for example, the FFs of  $\phi$ in the  process $e^+e^-\rightarrow \phi X$ by the  BESIII Collaboration~\cite{bes-phi}.
However, the theoretical relations of tensor-polarized FFs have not been completely investigated. In this work, we intend to derive the LIRs, QCD e.o.m., and WW-type relations  for the tensor-polarized FFs in a spin-1 hadron, which can provide constraints for the future experimental and theoretical studies of these FFs.

This paper is organized as follows. In Sec.~\ref{FFs-def}, we define the intrinsic, kinematical, and dynamical twist-3 FFs, and general properties of them are discussed.
We derive the theoretical relations among tensor-polarized FFs using QCD e.o.m. for quarks in Sec.~\ref{eom}.
The operator identities are obtained for the nonlocal quark-quark  and  quark-gluon-quark operators, then LIRs  and WW-type relations are also given based on the matrix elements of the operator identities in Sec.~\ref{lir}. A brief summary of this work is presented  in Sec.~\ref{summary}.

\section{Tensor-polarized fragmentation functions }
\label{FFs-def}
The tensor polarization is often indicated by the matrix $T$ for a spin-1 hadron, and the covariant form of $T^{\mu \nu}$  is
expressed as~\cite{Bacchetta:2000jk,Boer:2016xqr}
\begin{align}
T^{\mu\nu} & = \frac{1}{2} \left [ \frac{4}{3} S_{LL} \frac{(P_h^-)^2}{M^2}
                n^\mu n^\nu
          - \frac{2}{3} S_{LL} (  n^{\{ \mu} \bar n^{\nu \}} -g_T^{\mu\nu} )
              + \frac{1}{3} S_{LL} \frac{M^2}{(P_h^-)^2} \bar{n}^\mu \bar{n}^\nu
 + \frac{P_h^-}{M}  n^{\{ \mu} S_{LT}^{\nu \}}
- \frac{M}{2 P_h^-} \bar n^{\{ \mu} S_{LT}^{\nu \}}
+ S_{TT}^{\mu\nu} \right ],
\label{eqn:spin-1-tensor-1}
\\[-0.90cm] \nonumber
\end{align}
where $P_h$ and  $M$ are momentum and  mass for the produced hadron, respectively.
$a^{\left\{ \mu \right.} b^{ \left. \nu \right\} } =a^{\mu}b^{\nu}+a^{\nu}b^{\mu}$ denotes symmetrization of the indices.
The lightcone vectors $n$ and $\bar{n}$ are given by
\begin{align}
n^\mu =\frac{1}{\sqrt{2}} (\, 1,\, 0,\, 0,\,  -1 \, ), \ \
\bar n^\mu =\frac{1}{\sqrt{2}} (\, 1,\, 0,\, 0,\,  1 \, ),
\label{eqn:lightcone-n-nbar}
\end{align}
and $P_h$ can be written as $P_h=P_h^- n +\frac{M^2}{2 P_h^- } \bar{n}$.
For a Lorentz vector $a^{\mu}$, the lightcone components $a^{\pm}$ and transverse component $a_T$ are defined by
\begin{align}
a^{+}=a\cdot n, \, \,   a^{-}=a\cdot \bar{n}, \, \, a_T^{\mu}=g_T^{\mu \nu}a_{\nu}
\label{eqn:trans}
\end{align}
with
\begin{align}
g_T^{\mu \nu}=g^{\mu \nu}-n^{\mu}\bar{n}^{\nu}-n^{\nu}\bar{n}^{\mu}.
\label{eqn:trans-sub}
\end{align}
In Eq.~(\ref{eqn:spin-1-tensor-1}), $S_{LL}$, $S_{LT}^{\mu}$ and $S_{TT}^{\mu \nu}$ are the parameters which indicate different types of tensor polarization.

For a spin-1 hadron, the fragmentation correlator is defined as \cite{kumano:2021,Chen:2016moq, Bacchetta:2000jk,Ji:1993vw}
\begin{align}
\Delta_{ij} (z)=& \frac{1}{N_c}
\int  \! \frac{d \xi^+}{2\pi} \,   e^{i  \frac{P_h^-\xi^+}{z} }
\langle  0  \left | \mathcal W\left[\infty^+; \xi^+ \right]  q _i (\xi^+)   \right | P_h,T; X \rangle
\langle \, P_h, T; X   \left | \,
\bar{q}_j (0) \,  \mathcal W\left[0^+; \infty^+\right]  \right | 0 \rangle,  \nonumber \\
=& \frac{1}{z} \Big\{ S_{LL} \slashed{n}   F_{1LL}(z)+\frac{ M}{P_h^-}  \left[ \slashed{S}_{LT} F_{LT}(z)  +  S_{LL}   E_{LL}(z) \right]+ \sigma^{i+} S_{LT,i} H_{1LT}   \nonumber \\
&+  \frac{ M}{P_h^-} \left[  S_{LL}  \sigma^{-+}  H_{LL}(z)+ \gamma_5 \gamma_i \epsilon_T^{ij} S_{LT,j}  G_{LT}\right]   \Big\}
\label{eqn:ff1}
\end{align}
where $z$ is the longitudinal  momentum fraction carried by the produced hadron, $N_c$ is the number of color, and $\mathcal W$ is a Wilson line which ensures color gauge invariance. The transverse tensor $\epsilon_{T}^{\alpha \beta } $ is given by

\begin{align}
\epsilon_{T}^{\alpha \beta}=\epsilon^{\alpha \beta \mu \nu} n_{\mu} \bar{n}_{\nu}
\label{eqn:trasn-ten}
\end{align}
with the convention $\epsilon^{0123}=1$. In Eq.~(\ref{eqn:ff1}),  the correlator is expressed in terms of  six tensor-polarized FFs  up to twist 3, and the FFs  are real functions  with the  support region of $0<z<1$. $F_{1LL}(z)$ and $H_{1LT}(z)$ are leading-twist FFs, and the rest are also called intrinsic twist-3 FFs~\cite{Kanazawa:2015ajw}.
Since time-reversal invariance is not a necessary constraint for the fragmentation correlator, the last three FFs are actually  time-reversal odd FFs.
Note that there are also unpolarized and vector-polarized FFs in the correlator, which are neglected here since we are interested in the tensor-polarized ones.

The kinematical twist-3 FFs are related to the TMD FFs. In case of a tensor-polarized hadron, the TMD fragmentation correlator reads~\cite{Collins:1981uw,Collins:1992kk,Boer:2003bw,Bacchetta:2006tn,Metz:2016swz}
\begin{align}
\Delta_{ij} (z,k_T)=& \frac{1}{N_c}
\int  \! \frac{d \xi^+ d^2\xi_T}{(2\pi)^3} \,   e^{i (k^-\xi^++k_T\cdot \xi_T) }
\langle  0  \left | \mathcal W_1\left[\infty; \xi \right]  q _i (\xi)   \right | P_h,T; X \rangle
\langle \, P_h, T; X   \left | \,
\bar{q}_j (0) \,  \mathcal W_2\left[0; \infty\right]  \right | 0 \rangle_{\xi^-=0}
\label{eqn:tmd}
\end{align}
with
\begin{align}
\mathcal W_1\left[\infty; \xi \right]&=\mathcal W\left[\infty^+,\infty_T;\infty^+,\xi_T \right ] \mathcal W\left[\infty^+,\xi_T ;\xi^+,\xi_T \right], \nonumber \\
\mathcal W_2\left[0; \infty\right]&=\mathcal W\left[0^+,0_T; \infty^+,0_T \right]\mathcal W\left[\infty^+,0_T ;\infty^+,\infty_T \right ],
\label{eqn:tmd-wl}
\end{align}
and the correlator can be written in terms of  TMD FFs~\cite{Bacchetta:2000jk,Chen:2016moq,Chen:2023kqw}. The $k_T$-weighted FFs are defined with the help of the TMD fragmentation correlator,
\begin{align}
\Delta_{\partial,ij}^{\nu} (z)=\int d^2k_T k_T^{\nu} \Delta_{ij} (z,k_T),
\label{eqn:tmd-kine-ff}
\end{align}
which is parametrized by four $k_T$-weighted FFs at twist 3 \cite{Bacchetta:2000jk},
\begin{align}
\Delta_{\partial}^{\nu}(z)=\frac{M}{z} \left[ -S_{LT}^\nu  \slashed{n}  F_{1LT}^{(1)}(z) -    \epsilon_{T}^{\nu \rho} S_{LT \rho } \gamma_5  \slashed{n}  G_{1LT}^{(1)}(z)   +
S_{LL} \sigma^{\nu \alpha} n_{\alpha}  H_{1LL}^{(1)}(z)
-S_{TT}^{\nu \alpha} \sigma_{\alpha \beta}  n^{\beta} H_{1TT}^{(1)}(z)  \right],
\label{eqn:ffqkt1}
\end{align}
and these FFs are also called kinematical twist-3 FFs in Ref.~\cite{Kanazawa:2015ajw}. Due to Eq.~(\ref{eqn:tmd-kine-ff}), the kinematical twist-3 FFs are given by TMD FFs,
\begin{align}
F^{(1)}(z)=-z^2\int d^2k_T \frac{k_T^2}{2M^2} F(z,z^2k_T^2),
\label{eqn:tmd-kin}
\end{align}
where $F(z,z^2k_T^2)$ is a TMD FF.

Similarly, we define the collinear three-parton fragmentation correlator~\cite{Kumano:2021fem},
\begin{align}
\Delta^{\nu}_{F,ij} (z, z_1)=& \frac{1}{N_c}
\int  \! \frac{d \xi^+}{2\pi}  \frac{d \xi_1^+}{2\pi}    \,   e^{i P_h^-\xi^+ \frac{1}{z_1}  + i P_h^-\xi_1^+ (\frac{1}{z} - \frac{1}{z_1} ) }
\langle  0  \left | \mathcal W\left[\infty^+; \xi_1^+ \right] igF^{- \nu}(\xi_1^+) \mathcal W\left[\xi_1^+; \xi^+ \right]  q _i (\xi^+)   \right | P_h,T; X \rangle \nonumber \\
& \times \langle \, P_h,T; X   \left | \,
\bar q _j (0) \mathcal W\left[0^+; \infty^+\right] \right | 0 \rangle.
\label{eqn:ffqg1}
\end{align}
By  inserting a complete set of intermediate states, one can prove that
\begin{align}
\Delta^{\nu}_{F} (z, z)=0, \quad \Delta^{\nu}_{F} (z, 0)=0,
\label{eqn:ffsp}
\end{align}
and this corresponds to the vanishing partonic
pole matrix elements which are important to understand the SSAs in the hard semi-inclusive processes~\cite{Meissner:2008yf}.
Then, the support region of $\Delta^{\nu}_{F} (z, z_1)$  is
\begin{align}
0 \leq z \leq 1, \quad  0 < \frac{z}{z_1} < 1.
\label{eqn:ffspa}
\end{align}
Taking the derivative of this correlator with
respect to $1/z_1$ and then setting $z_1=z$, one can also obtain \cite{Kanazawa:2015ajw}
\begin{align}
\frac{\partial \Delta^{\nu}_{F} (z, z_1)}{\partial (1/z_1)}|_{z_1=z}=0.
\label{eqn:ffsp1}
\end{align}
The parametrization of $\Delta^{\nu}_{F} (z, z_1)$ is just a copy of the corresponding
three-parton distribution correlator~\cite{Kumano:2021fem}, and it can be expressed in terms of four dynamical FFs at twist 3,
\begin{align}
\Delta^{\nu}_{F,ij} (z, z_1)= \frac{M}{z} \left[ -S_{LT}^\nu  \slashed{n}  \hat{F}_{LT}(z, z_1) -    i \epsilon_{T}^{\nu \rho} S_{LT \rho } \gamma_5  \slashed{n}  \hat{G}_{LT}(z, z_1)   -S_{LL} \gamma^{\nu}  \slashed{n}  \hat{H}_{LL}^{\bot}(z, z_1) -S_{TT}^{\nu \rho} \gamma_{\rho}  \slashed{n} \hat{H}_{TT}(z, z_1)  \right].
\label{eqn:ffqg2}
\end{align}
Note that the dynamical FFs are complex functions which are different from the intrinsic and kinematical ones.

\section{Equation of motion relations for FFs }
\label{eom}
The intrinsic, kinematical and dynamical FFs are not independent functions, since they can be related to each other by the e.o.m. relations. For a spin-1/2 hadron, the e.o.m. relations for FFs were derived in Refs.~\cite{Kanazawa:2015ajw,Metz:2012ct} based on the QCD e.o.m. for quarks, namely, $(i\slashed{D}-m_q )q(x)=0$. In the following, we will investigate the e.o.m. relations for tensor-polarized FFs.   After some algebra, the QCD e.o.m. for quarks becomes
\begin{align}
(iD^{\mu}+\sigma^{\mu \nu} D_{\nu} +m_q \gamma^{\mu})q(x)=0,
\label{eqn:eomffa1}
\end{align}
where $m_q$ is the mass of the quark.
If we set $\mu=-$ and take the corresponding  matrix element for Eq.~(\ref{eqn:eomffa1}), an e.o.m. relation can be obtained for the intrinsic, kinematical and dynamical FFs,
\begin{align}
\frac{E_{LL}(z)}{z}+ \frac{i H_{LL}(z)}{z}-\frac{m_q}{M}F_{1LL}(z)=2 \left[  -i H^{(1)}_{1LL}(z)+ \mathcal{P}\int^{\infty}_{z} \frac{dz_1}{(z_1)^2} \frac{\hat{H}_{LL}^{\bot}(z,z_1)}{\frac{1}{z}-\frac{1}{z_1}} \right],
\label{eqn:eom1}
\end{align}
where $\mathcal{P}$ stands for the principal integral, and it can be neglected due to Eq.~(\ref{eqn:ffsp}). All the FFs  in Eq.~(\ref{eqn:eom1}) are  related to the $S_{LL}$-type tensor polarization. Furthermore, this relation can be reexpressed in terms of the real and complex parts of the dynamical FF,
\begin{align}
\frac{E_{LL}(z)}{z}=&2 \int^{\infty}_{z} \frac{dz_1}{(z_1)^2} \frac{ \mathrm{Re}\left[   \hat{H}_{LL}^{\bot}(z, z_1) \right]}{ \frac{1  }{z}- \frac{1}{z_1}  }  +\frac{m_q}{M}F_{1LL}(z), \label{eqn:eomffa2a}\\
 \frac{ H_{LL}(z)}{z}=&2 \int^{\infty}_{z} \frac{dz_1}{(z_1)^2}   \frac{ \mathrm{Im}\left[  \hat{H}_{LL}^{\bot}(z, z_1) \right]}{ \frac{1  }{z}- \frac{1}{z_1}  }  -2 H_{1LL}^{(1)}(z).
\label{eqn:eom1a}
\end{align}
We can see that the time-reversal even and odd  intrinsic FFs are related to the real and imaginary parts of the dynamical FFs, respectively.
If we neglect the quark mass, the intrinsic twist-3 FFs $E_{LL}(z)$ and $H_{LL}(z)$ are given by the kinematical and dynamical twist-3 FFs.

Multiplying $\gamma^{\nu}$ on the l.h.s. of Eq.~(\ref{eqn:eomffa1}), then antisymmetrizing  $\mu$ and $\nu$,
we can obtain the identity as
\begin{align}
\left[i ( \gamma^{\mu} D^{\nu}- \gamma^{\nu} D^{\mu} )-  \epsilon^{\mu \nu \rho \sigma} \gamma_{\sigma} \gamma_{5} D_{\rho} +im_q \sigma^{\mu \nu}  \right]q(x)=0.
\label{eqn:eom2}
\end{align}
Analogously, we set $\mu=-$ and consider $\nu$ as a transverse component, then the matrix element of Eq.~(\ref{eqn:eom2}) leads to
\begin{align}
&\frac{F_{LT}(z)}{z}+ \frac{i G_{LT}(z)}{z} +\frac{im_q}{M}H_{1LT}(z) \nonumber \\
=& -iG_{1LT}^{(1)}(z)+ \int^{\infty}_{z} \frac{dz_1}{(z_1)^2} \frac{\hat{G}_{LT}(z,z_1)}{\frac{1}{z}-\frac{1}{z_1}} - \left[ F_{1LT}^{(1)}(z) +\int^{\infty}_{z} \frac{dz_1}{(z_1)^2} \frac{\hat{F}_{LT}(z,z_1)}{\frac{1}{z}-\frac{1}{z_1}} \right ],
\label{eqn:eom2a}
\end{align}
and it indicates the relation among the intrinsic, kinematical and dynamical FFs for the $S_{LT}$-type tensor polarization.
Furthermore, Eq.~(\ref{eqn:eom2a}) can be divided into two identities,
\begin{align}
\frac{F_{LT}(z)}{z}=&-  \int_{z}^{\infty} \frac{dz_1}{(z_1)^2}      \frac{ \mathrm{Re} \left[  \hat{F}_{LT}(z,z_1)-\hat{G}_{LT}(z, z_1) \right]}{ \frac{1  }{z}- \frac{1}{z_1}  }    -  F^{(1)}_{1LT}(z), \label{eqn:eom2b1}  \\
\frac{G_{LT}(z)}{z}=&-\int_{z}^{\infty} \frac{dz_1}{(z_1)^2}   \frac{ \mathrm{Im} \left[  \hat{F}_{LT}(z,z_1)-  \hat{G}_{LT}(z, z_1) \right]}{ \frac{1  }{z} - \frac{1}{z_1}  }- G^{(1)}_{1LT}(z) -  \frac{m_q}{M}H_{1LT}(z).
\label{eqn:eom2b}
\end{align}

As indicated by Eq.~(\ref{eqn:ff1}), there are no intrinsic FFs for the $S_{TT}$-type tensor polarization. However, we can also derive the following identity using  Eq.~(\ref{eqn:eom2}):
\begin{align}
i H^{(1)}_{1TT}(z)+\int^{\infty}_{z} \frac{dz_1}{(z_1)^2} \frac{\hat{H}_{TT}(z,z_1)}{\frac{1}{z}-\frac{1}{z_1}} =0,
\label{eqn:eom3a}
\end{align}
and it implies
\begin{align}
\int^{\infty}_{z} \frac{dz_1}{(z_1)^2} \frac{\mathrm{Re}\left[ \hat{H}_{TT}(z,z_1)\right]}{\frac{1}{z}-\frac{1}{z_1}} =&0, \\
 H^{(1)}_{1TT}(z)+\int^{\infty}_{z} \frac{dz_1}{(z_1)^2} \frac{\mathrm{Im}\left[\hat{H}_{TT}(z,z_1)\right]}{\frac{1}{z}-\frac{1}{z_1}} =&0,
\label{eqn:eom3b}
\end{align}
which complete the derivation of the QCD e.o.m. relations for tensor-polarized FFs.

\section{Lorentz invariance and  Wandzura-Wilczek-type relations }
\label{lir}
Taking the derivative of nonlocal quark-quark operators, one can obtain the identities where the quark-quark operators are expressed in terms of the quark-gluon-quark ones.
The theoretical relations have been investigated for PDFs, FFs and distribution amplitudes  by using these
 identities of nonlocal operators, and this method was well explained in Refs.~\cite{Kanazawa:2015ajw,Balitsky:1987bk,Balitsky:1989ry,Balitsky:1990ck,Ball:2001uk,Ball:1998ff,Ball:1996tb,Braun:1989iv,Kodaira:1998jn,Eguchi:2006qz}. In this section, we adopt the same method to derive the theoretical relations for twist-3 tensor-polarized FFs such as LIRs and WW-type relations. We first consider the derivative of the nonlocal quark-quark operator~\cite{Kanazawa:2015ajw},
\begin{align}
&\frac{\partial}{\partial \xi_{\alpha}}
\langle  0   |  \mathcal {W}\left[\infty \xi; -\xi \right]  q (-\xi)  | P_h,T; X \rangle
\langle \, P_h, T; X   | \,
\overline{q}(\xi)  \Gamma \mathcal W\left[\xi; \infty \xi \right]   | 0 \rangle  \nonumber \\
=&- \langle  0   | \mathcal W\left[\infty \xi; -\xi \right] \overset{\rightarrow}{D}\!\!\!\!\!\phantom{D}^{\alpha}(-\xi)  q (-\xi)    | P_h,T; X \rangle \langle \, P_h, T; X    | \,
\overline{q}(\xi)  \Gamma \mathcal W\left[\xi; \infty \xi \right]   | 0 \rangle  \nonumber \\
&+\langle  0  \left | \mathcal W\left[\infty \xi; -\xi \right] q (-\xi)   \right | P_h,T; X \rangle
\langle \, P_h, T; X   | \,
\overline{q}(\xi) \overset{\leftarrow}{D}\!\!\!\!\!\phantom{D}^{\alpha}(\xi)  \Gamma \mathcal W\left[\xi; \infty \xi \right] | 0 \rangle    \nonumber \\
&+ i \int^{\infty}_{-1} dt t \langle  0  \left | \mathcal W\left[\infty \xi; t \xi \right]
gF^{\alpha \xi}(t \xi) \mathcal W\left[t \xi; -\xi \right] q (-\xi)   \right | P_h,T; X \rangle \langle \, P_h, T; X  | \,
\overline{q}(\xi)  \Gamma \mathcal W\left[\xi; \infty \xi \right]  | 0 \rangle
\nonumber \\
&+ i \int_{\infty}^{1} dt t
\langle  0 | \mathcal W\left[\infty \xi; -\xi \right]
q (-\xi)   | P_h,T; X \rangle \langle \, P_h, T; X   \left | \,
\overline{q}(\xi) \Gamma   \mathcal W\left[\xi; t \xi \right]   gF^{\alpha \xi}(t \xi)     \mathcal W\left[t\xi; \infty \xi \right]  \right | 0 \rangle,
\label{ope-d}
\end{align}
where $\xi$ is not necessarily a lightcone vector and $\Gamma$ is a gamma matrix. In Eq.~(\ref{ope-d}), the terms with the covariant derivative $D^{\alpha}$ can be replaced by the total derivative of the nonlocal quark-quark operator, which is  related to the translation of the operator, and its matrix element can be expressed as~\cite{Kanazawa:2015ajw}
\begin{align}
&\bar{\partial}^\rho \langle  0  \left |  \mathcal W\left[\infty \xi; -\xi \right]  q (-\xi)   \right | P_h,T; X \rangle
\langle \, P_h, T; X   \left | \,
\overline{q}(\xi)  \Gamma_1 \mathcal W\left[\xi; \infty \xi \right]  \right | 0 \rangle  \nonumber \\
=&\lim_{x_{\rho}\rightarrow 0 }\frac{d}{dx_{\rho}} \langle  0  \left |  \mathcal W\left[\infty \xi+x; -\xi+x \right]  q (-\xi+x)   \right | P_h,T; X \rangle
\langle \, P_h, T; X   \left | \,
\overline{q}(\xi+x)  \Gamma_1 \mathcal W\left[\xi+x; \infty \xi+x \right]  \right | 0 \rangle
\nonumber \\
=&
\langle  0  | \mathcal W\left[\infty \xi; -\xi \right] \overset{\rightarrow}{D}\!\!\!\!\!\phantom{D}^{\rho}(-\xi)  q (-\xi)  | P_h,T; X \rangle \langle \, P_h, T; X    | \,
\bar{q}(\xi)  \Gamma_1 \mathcal W\left[\xi; \infty \xi \right]  | 0 \rangle  \nonumber \\
&+\langle  0  \left | \mathcal W\left[\infty \xi; -\xi \right] q (-\xi)   \right | P_h,T; X \rangle
\langle \, P_h, T; X   | \,
\bar{q}(\xi) \overset{\leftarrow}{D}\!\!\!\!\!\phantom{D}^{\rho}(\xi)  \Gamma_1 \mathcal W\left[\xi; \infty \xi \right]  | 0 \rangle    \nonumber \\
&+ \int^{\infty}_{-1} dt \langle  0  \left | \mathcal W\left[\infty \xi; t \xi \right]i
gF^{\rho \xi}(t \xi) \mathcal W\left[t \xi; -\xi \right] q (-\xi)   \right | P_h,T; X \rangle \langle \, P_h, T; X   \left | \,
\bar{q}(\xi)  \Gamma_1 \mathcal W\left[\xi; \infty \xi \right]  \right | 0 \rangle
\nonumber \\
&+ \int_{\infty}^{1} dt
\langle  0  \left | \mathcal W\left[\infty \xi; -\xi \right]
q (-\xi)   \right | P_h,T; X \rangle \langle \, P_h, T; X   \left | \,
\bar{q}(\xi) \Gamma_1   \mathcal W\left[\xi; t \xi \right] i  gF^{\rho \xi}(t \xi)     \mathcal W\left[t\xi; \infty \xi \right]  \right | 0 \rangle,
\label{ope-tot}
\end{align}
where $\Gamma_1$ stands for a gamma matrix such as $\gamma^{\mu}$ and $\sigma^{\mu \nu}$. Due to the translation invariance, the matrix element in Eq.~(\ref{ope-tot}) should vanish.

In the following, the Wilson lines are neglected in the operator identities, since this will not cause confusion. We derive a relation between quark-quark and quark-gluon-quark operators  by choosing $\Gamma=(g^{\rho \alpha}g^{\lambda}_{\ \, \sigma}-g^{\alpha}_{\ \, \sigma}g^{\rho \lambda})\gamma_{\lambda}$ in Eq.~(\ref{ope-d}) and $\Gamma_1=(\sigma^{\sigma \beta}\gamma^{\rho}+\gamma^{\rho}\sigma^{\sigma \beta})$ in Eq.~(\ref{ope-tot}),
\begin{align}
&\xi_{\alpha} \left[ \frac{\partial}{\partial \xi_{\alpha}}
\langle  0   |   q (-\xi)  | P_h,T; X \rangle
\langle \, P_h, T; X   | \,
\overline{q}(\xi)  \gamma^{\sigma}    | 0 \rangle -\frac{\partial}{\partial \xi_{\sigma}}
\langle  0   |   q (-\xi)  | P_h,T; X \rangle
\langle \, P_h, T; X   | \,
\overline{q}(\xi)  \gamma^{\alpha}    | 0 \rangle \right ]
\nonumber \\
=& \left[ \int^{\infty}_{-1} dt \langle  0  \left |
gF_{\rho \xi}(t \xi)  q (-\xi)   \right | P_h,T; X \rangle \langle \, P_h, T; X  | \,
\overline{q}(\xi)  \gamma_{\tau} \gamma_5| 0 \rangle
+ \int_{\infty}^{1} dt
\langle  0 |
q (-\xi)   | P_h,T; X \rangle \langle \, P_h, T; X   \left | \,
\overline{q}(\xi) \gamma_{\tau}   \gamma_5    gF_{\rho \xi}(t \xi)       \right | 0 \rangle \right ]
\nonumber \\
& \times \epsilon^{\sigma \xi \rho \tau} - i \int^{\infty}_{-1} dt t \langle  0  \left |
gF^{\sigma \xi}(t \xi)  q (-\xi)   \right | P_h,T; X \rangle \langle \, P_h, T; X  | \,
\overline{q}(\xi)  \slashed{\xi}  | 0 \rangle
- i \int_{\infty}^{1} dt t
\langle  0 |q(-\xi)   | P_h,T; X \rangle  \nonumber \\
& \times \langle \, P_h, T; X   \left | \,
\overline{q}(\xi) \slashed{\xi}      gF^{\sigma \xi}(t \xi)       \right | 0 \rangle,
\label{ope-vec}
\end{align}
where the matrix element of the total derivative operator is neglected, and the quark mass terms vanish. The quark-quark operator appears in the l.h.s. of Eq.~(\ref{ope-vec}), which can be written in terms of the intrinsic tensor-polarized FFs as shown in  Eq.~(\ref{eqn:ff1}).
If the vector $\xi$ is not necessarily on the lightcone, the matrix element of the nonlocal quark-quark operator can be expressed as
\begin{align}
&\langle  0   |   q (-\xi)  | P_h,T; X \rangle
\langle \, P_h, T; X   | \,
\overline{q}(\xi)  \gamma^{\sigma}    | 0 \rangle \nonumber \\
=&8N_cM^2 \int d\left(\frac{1}{z}\right) e^{\frac{2iP\xi}{z}} \left[  \frac{3}{4}A^{\sigma} \frac{F_{1LL}(z)}{z}   + B^{\sigma}   \frac{F_{LT}(z)}{z}     \right]
\label{ope-vec-le}
\end{align}
with
\begin{align}
A^{\sigma}=& \frac{\xi \cdot T\cdot \xi}{(P_h \cdot \xi)^2} P_h^{\sigma}-M^2 \frac{\xi \cdot T\cdot \xi}{(P_h \cdot \xi)^3} \xi^{\sigma},  \\
B^{\sigma}=& \frac{ T^{\sigma \mu}  \xi_{\mu}}{P_h \cdot \xi}-\frac{\xi \cdot T\cdot \xi}{(P_h \cdot \xi)^2} P_h^{\sigma}+M^2 \frac{\xi \cdot T\cdot \xi}{(P_h \cdot \xi)^3} \xi^{\sigma}.
\label{ope-vec-le1}
\end{align}
Eq.~(\ref{ope-vec-le}) is exact at twist 3 since the twist-4 FFs are not included.
We substitute Eq.~(\ref{ope-vec-le}) into Eq.~(\ref{ope-vec}) to estimate the derivative, and take the lightcone limt of $\xi^2 \rightarrow 0$,  then, the l.h.s. of Eq.~(\ref{ope-vec}) is given by
$F_{1LL}(z)$ and $F_{LT}(z)$. The r.h.s. of  Eq.~(\ref{ope-vec}) can be directly calculated with the help of Eq.~(\ref{eqn:ffqg2}), and we obtain the following relation by combining the l.h.s. and r.h.s.,
\begin{align}
&\frac{3}{2} \tilde{F}_{1LL}(z)+\frac{1}{z} \frac{d\tilde{F}_{LT}(z)}{d(1/z)} \nonumber \\
=&  \int d\left(\frac{1}{z_1}\right) \mathcal{P} \left( \frac{1}{\frac{1}{z}-\frac{1}{z_1}} \right)  \left \{ \Big(\frac{\partial }{\partial (1/z)}+ \frac{\partial }{\partial(1/z_1)} \Big) \mathrm{Re} \left[ \tilde{G}_{LT}(z,z_1)\right ] - \Big(\frac{\partial }{\partial (1/z)}- \frac{\partial }{\partial (1/z_1)} \Big) \mathrm{Re}\left [ \tilde{F}_{LT}(z,z_1) \right]  \right \},
\label{j-ope}
\end{align}
where the convention of  $\tilde{F}(z)=F(z)/z$ is used for  a intrinsic or kinematical FF $F(z)$,  and $\tilde{F}(z,z_1)=\hat{F}(z,z_1)/z$ for a dynamical one.
Combining Eq.~(\ref{j-ope}) with the e.o.m relation of  Eq.~(\ref{eqn:eom2b1}), one can obtain
\begin{align}
\frac{3}{2}\tilde{F}_{1LL}(z)-\tilde{F}_{LT}(z)  - (1-z \frac{d}{dz})F_{1LT}^{(1)}(z)
=  -2 \int_z^{\infty} \frac{dz_1}{(z_1)^2}  \frac{\mathrm{Re} \left [ \tilde{F}_{LT}(z,z_1) \right] }{ (\frac{1}{z}-\frac{1}{z_1})^2},
\label{lir1}
\end{align}
and this is a new LIR for tensor-polarized FFs. If we integrate Eq.~(\ref{j-ope}) over the momentum fraction $z$, one can have
\begin{align}
F_{LT}(z)= & - \frac{3z}{2} \int_z^1 d z_1 \frac{F_{1LL}\left(z_1 \right)}{(z_1)^2}+z \int_z^1 \frac{d z_1}{z_1} \int_{z_1}^{\infty} \frac{d z_2}{(z_2)^2} \Bigg \{ \frac{\left[1+\frac{1}{z_1} \delta(\frac{1}{z_1}-\frac{1}{z})\right] \mathrm{Re}\left[\hat{G}_{LT}(z_1, z_2)\right]}{\frac{1}{z_1}-\frac{1}{z_2}} \nonumber \\
&-\frac{\left[\frac{3}{z_1}-\frac{1}{z_2}+\frac{1}{z_1}(\frac{1}{z_1}-\frac{1}{z_2}) \delta(\frac{1}{z_1}-\frac{1}{z})\right] \mathrm{Re}\left[\hat{F}_{LT}\left(z_1, z_2\right)\right]}{(\frac{1}{z_1}-\frac{1}{z_2})^2} \Bigg \},
\label{ww1a}
\end{align}
where it should be understood that $z_1$ falls within  the range of integration $(z,1)$, namely, $\int_z^1 dz_1 F(z_1)\delta(1/z_1-1/z)=z^2F(z)$.
The intrinsic twist-3 FF $F_{LT}(z)$ is decomposed into the contributions of a twist-2 FF $F_{1LL}$ and the dynamical FFs. We can obtain  a similar expression for the kinematical twist-3 FF $F_{1LT}^{(1)}(z)$ by inserting Eq.~(\ref{ww1a}) into the e.o.m. relation of Eq.~(\ref{eqn:eom2b1}),
\begin{align}
F_{1LT}^{(1)}(z)=\frac{3}{2} \int_z^1 d z_1 \frac{F_{1LL}\left(z_1 \right)}{(z_1)^2}+ \int_z^1 \frac{d z_1}{z_1} \int_{z_1}^{\infty} \frac{d z_2}{(z_2)^2} \Bigg \{
\frac{\left(\frac{3}{z_1}-\frac{1}{z_2}\right) \mathrm{Re}\left[\hat{F}_{LT}\left(z_1, z_2\right)\right]}{\left(\frac{1}{z_1}-\frac{1}{z_2}\right)^2}-
\frac{ \mathrm{Re}\left[\hat{G}_{LT}\left(z_1, z_2\right)\right]}{\frac{1}{z_1}-\frac{1}{z_2}}  \Bigg \}.
\label{ww1b}
\end{align}
By dropping the contributions of dynamical FFs into Eqs.~(\ref{ww1a}) and (\ref{ww1b}), they become the WW-type relations. Then, the twist-3 intrinsic and kinematical FFs
can be estimated by using the twist-2 FF $F_{1LL}(z)$, and the latter should be much easier to be extracted from experimental measurements compared with the former.

If we choose $\Gamma=\sigma^{\mu \alpha}$ in Eq.~(\ref{ope-d}) and $\Gamma_1=1$ in Eq.~(\ref{ope-tot}), the following identity can be derived~\cite{Kanazawa:2015ajw},
\begin{align}
&\frac{\partial}{\partial \xi_{\alpha}}
\langle  0   |   q (-\xi)  | P_h,T; X \rangle
\langle \, P_h, T; X   | \,
\overline{q}(\xi)  \sigma^{\xi \alpha}    | 0 \rangle
\nonumber \\
=&  \int^{\infty}_{-1} dt t \langle  0  \left |
igF_{\alpha \xi}(t \xi)  q (-\xi)   \right | P_h,T; X \rangle \langle \, P_h, T; X  | \,
\overline{q}(\xi)  \sigma^{\xi \alpha} | 0 \rangle  \nonumber \\
&+\int_{\infty}^{1} dt t
\langle  0 |
q (-\xi)   | P_h,T; X \rangle \langle \, P_h, T; X   \left | \,
\overline{q}(\xi) \sigma^{\xi \alpha}       igF_{\alpha \xi}(t \xi)       \right | 0 \rangle.
\label{ope-ten}
\end{align}
Similarly, the matrix element of the nonlocal operator  $\overline{q}(\xi)  \sigma^{\xi \sigma} q (-\xi)$ is expressed in terms of the FFs $\tilde{H}_{1LT}(z)$ and $\tilde{H}_{LT}(z)$ at twist 3,
\begin{align}
&\langle  0   |   q (-\xi)  | P_h,T; X \rangle
\langle \, P_h, T; X   | \,
\overline{q}(\xi)  \sigma^{\xi \sigma}    | 0 \rangle \nonumber \\
=&-4N_cM \int d\left(\frac{1}{z}\right) e^{\frac{2iP\xi}{z}} \left[  2 (W^{\sigma}+V^{\sigma}) \tilde{H}_{1LT}(z)   + \frac{3}{2} V^{\sigma}   \tilde{H}_{LL}(z)    \right],
\label{ope-ten-ff}
\end{align}
where  $W^{\sigma}$ and $V^{\sigma}$ are defined as
\begin{align}
W^{\sigma}=& T^{\sigma \mu}  \xi_{\mu} - \frac{\xi \cdot T\cdot \xi}{P_h \cdot \xi}  P_h^{\sigma},  \\
V^{\sigma}=& M^2 \frac{\xi \cdot T\cdot \xi}{(P_h \cdot \xi)^2} \left[ \xi^{\sigma}- \frac{\xi^2}{P_h \cdot \xi}  P_h^{\sigma}\right],
\label{ope-ten-wv}
\end{align}
and they satisfy the relations of $W\cdot \xi=0$ and $V\cdot \xi=0$. In the lightcone limit $\xi^2 \rightarrow 0$, Eq.~(\ref{ope-ten-ff}) goes back to Eq.~(\ref{eqn:ff1}).
We obtain the following identity by calculating the matrix element in Eq.~(\ref{ope-ten}):
\begin{align}
4 \tilde{H}_{1LT}(z)-\frac{1}{z^2} \frac{dH_{LL}(z)}{d(1/z)}
= -2 \int d\left(\frac{1}{z_1}\right) \mathcal{P} \left( \frac{1}{\frac{1}{z}-\frac{1}{z_1}} \right)  \Big(\frac{\partial }{\partial (1/z)}- \frac{\partial }{\partial(1/z_1)} \Big) \mathrm{Im} \left[ \tilde{H}_{LL}^{\bot}(z,z_1)\right ].
\label{i-ope}
\end{align}
Moreover, we obtain $d(\tilde{H}_{LL}(z)/z)/d(1/z)$ by using the expression in Eq.~(\ref{eqn:eom1a}), and the sum of $d(\tilde{H}_{LL}(z)/z)/d(1/z)$ and Eq.~(\ref{i-ope}) leads to
\begin{align}
\tilde{H}_{LL}(z)+2\tilde{H}_{1LT}(z)  + (1-z \frac{d}{dz})H_{1LL}^{(1)}(z)
=  -2  \int_z^{\infty} \frac{dz_1}{(z_1)^2}  \frac{\mathrm{Im} \left [ \tilde{H}_{LL}^{\bot}(z,z_1) \right] }{ (\frac{1}{z}-\frac{1}{z_1})^2},
\label{lir2}
\end{align}
which is also a LIR for tensor-polarized FFs. The integration of Eq.~(\ref{i-ope}) gives
\begin{align}
H_{LL}(z)= & 4 \int_z^1 d z_1 \frac{H_{1LT}\left(z_1 \right)}{z_1}+4 \int_z^1 d z_1 \int_{z_1}^{\infty} \frac{d z_2}{(z_2)^2}
\frac{\frac{2}{z_1}-\frac{1}{z_2}+\frac{1}{2z_1}(\frac{1}{z_1}-\frac{1}{z_2}) \delta(\frac{1}{z_1}-\frac{1}{z}) }{(\frac{1}{z_1}-\frac{1}{z_2})^2}
\mathrm{Im}\left[\hat{H}_{LL}^{\bot}\left(z_1, z_2\right)\right],
\label{ww2a}
\end{align}
and the intrinsic  twist-3 FF $H_{LL}(z)$ is expressed in terms of the twist-2 FF $H_{1LT}(z)$ and the dynamical FF $\hat{H}_{LL}^{\bot}\left(z_1, z_2\right)$.
If we combine Eq.~(\ref{ww2a}) with the e.o.m. relation of Eq.~(\ref{eqn:eom1}),
\begin{align}
H_{1LL}^{(1)}(z)= &- \frac{2}{z} \int_z^1 d z_1 \frac{H_{1LT}\left(z_1 \right)}{z_1}-\frac{2}{z}\int_z^1 d z_1 \int_{z_1}^{\infty} \frac{d z_2}{(z_2)^2}
\frac{\frac{2}{z_1}-\frac{1}{z_2} }{(\frac{1}{z_1}-\frac{1}{z_2})^2}
\mathrm{Im}\left[\hat{H}_{LL}^{\bot}\left(z_1, z_2\right)\right],
\label{ww2b}
\end{align}
which also decomposes the kinematical twist-3  FF $H_{1LL}^{(1)}(z)$ into  the contributions of $H_{1LT}(z)$ and $\hat{H}_{LL}^{\bot}\left(z_1, z_2\right)$. We can obtain the WW-type relations for
$H_{LL}(z)$ and $H_{1LL}^{(1)}(z)$ by dropping the terms of the dynamical FF in Eqs.~(\ref{ww2a}) and (\ref{ww2b}).

If we consider the matrix elements of Eqs.~(\ref{ope-d}) and (\ref{ope-tot}) with $\Gamma=\epsilon^{\alpha \mu \rho S_{LT}} \gamma_{\mu} \gamma_5 $ and $\Gamma_1=\frac{i}{2}(\gamma^{\rho}\sigma^{S_{LT}\xi}- \sigma^{S_{LT}\xi}\gamma^{\rho})$, respectively, one can derive
\begin{align}
&  \epsilon^{\alpha \mu  \rho S_{LT}}  \xi_{\rho}     \frac{\partial}{\partial \xi^{\alpha}}
\langle  0   |   q (-\xi)  | P_h,T; X \rangle
\langle \, P_h, T; X   | \,
\overline{q}(\xi)  \gamma_{\mu}  \gamma_5  | 0 \rangle
\nonumber \\
=&  \int^{\infty}_{-1} dt \langle  0  \left |
gF_{\xi S_{LT}}(t \xi)  q (-\xi)   \right | P_h,T; X \rangle \langle \, P_h, T; X  | \,
\overline{q}(\xi)  \slashed{\xi} | 0 \rangle
+ \int_{\infty}^{1} dt
\langle  0 |
q (-\xi)   | P_h,T; X \rangle \langle \, P_h, T; X   \left | \,
\overline{q}(\xi) \slashed{\xi}       gF_{\xi S_{LT}}(t \xi)       \right | 0 \rangle
\nonumber \\
& +  i\epsilon^{\alpha \mu \xi S_{LT}}  \Big[   \int^{\infty}_{-1} dt t \langle  0  \left |
gF_{\alpha \xi}(t \xi)  q (-\xi)   \right | P_h,T; X \rangle \langle \, P_h, T; X  | \,
\overline{q}(\xi)  \gamma_{\mu} \gamma_5  | 0 \rangle
+\int_{\infty}^{1} dt t
\langle  0 |q(-\xi)   | P_h,T; X \rangle   \nonumber \\
&  \times \langle \, P_h, T; X   \left | \,
\overline{q}(\xi) \gamma_{\mu}  \gamma_5   gF_{\alpha \xi}(t \xi)       \right | 0 \rangle \Big ]  -2im_q \langle  0   |   q (-\xi)  | P_h,T; X \rangle
\langle \, P_h, T; X   | \,
\overline{q}(\xi)  \sigma^{\xi S_{LT}}  | 0 \rangle,
\label{ope-axial}
\end{align}
and the l.h.s. is related to the matric element of  $\overline{q}(\xi)  \gamma^{\mu} \gamma_5 q (-\xi)$, which is given by
\begin{align}
&\langle  0   |   q (-\xi)  | P_h,T; X \rangle
\langle \, P_h, T; X   | \,
\overline{q}(\xi)  \gamma^{\mu} \gamma_5   | 0 \rangle
=4N_cM   \frac{\epsilon^{\mu \xi \alpha P_h}}{P_h \cdot \xi} Y_{\alpha}    \int d\left(\frac{1}{z}\right) e^{\frac{2iP\xi}{z}}   \tilde{G}_{LT}(z),
\label{ope-axi-ff}
\end{align}
and the vector $Y$ is defined as
\begin{align}
Y^{\alpha}=\frac{2M}{P_h \cdot \xi} \left[ T^{\alpha \mu}  \xi_{\mu}-\frac{\xi \cdot T\cdot \xi}{P_h \cdot \xi}  P_h^{\alpha}
+  M^2 \frac{\xi \cdot T\cdot \xi}{(P_h \cdot \xi)^2}  (\xi^{\alpha}- \frac{\xi^2}{P_h \cdot \xi}  P_h^{\alpha})   \right].
\label{ope-axi-y}
\end{align}
If we take the lightcone limit $\xi^2\rightarrow 0$, one can obtain $Y^{\alpha} \rightarrow S_{LT}^{\alpha}$. Thus, Eq.~(\ref{ope-axial}) leads to the following identity,
\begin{align}
&\frac{1}{z} \frac{d\tilde{G}_{LT}(z)}{d(1/z)} + \frac{m_q}{M} \frac{d\tilde{H}_{1LT}(z)}{d(1/z)}  \nonumber \\
=&  \int d\left(\frac{1}{z_1}\right) \mathcal{P} \left( \frac{1}{\frac{1}{z}-\frac{1}{z_1}} \right)  \left \{ \Big(\frac{\partial }{\partial (1/z)}-\frac{\partial }{\partial (1/z_1)} \Big) \mathrm{Im}\left [ \tilde{G}_{LT}(z,z_1) \right]-
\left(\frac{\partial }{\partial (1/z)}+ \frac{\partial }{\partial(1/z_1)} \right) \mathrm{Im} \left[ \tilde{F}_{LT}(z,z_1)\right ]   \right \}.
\label{j-axi}
\end{align}
Combining Eq.~(\ref{j-axi}) with the e.o.m. relation of Eq.~(\ref{eqn:eom2b}), another LIR can be derived for tensor-polarized FFs,
\begin{align}
\tilde{G}_{LT}(z)+ (1-z \frac{d}{dz})G_{1LT}^{(1)}(z)
=  -2   \int_z^{\infty} \frac{dz_1}{(z_1)^2}   \frac{\mathrm{Im} \left [ \tilde{G}_{LT}(z,z_1) \right] }{ (\frac{1}{z}-\frac{1}{z_1})^2}   .
\label{lir3}
\end{align}
and the quark mass term in Eq.~(\ref{j-axi}) is canceled in this LIR. From Eqs.~(\ref{j-axi}) and (\ref{eqn:eom2b}), one can also express the twist-3 FFs $G_{LT}(z)$ and $G_{1LT}^{(1)}(z)$ in terms of
$H_{1LT}(z)$, $\hat{F}_{LT}\left(z_1, z_2\right)$ and  $\hat{G}_{LT}\left(z_1, z_2\right)$,
\begin{align}
G_{LT}(z)= & -\frac{m_q}{M} \left[ zH_{1LT}(z)+ z \int_z^1 d z_1 \frac{H_{1LT}(z_1 )}{z_1} \right]  -z \int_z^1 \frac{d z_1}{z_1} \int_{z_1}^{\infty} \frac{d z_2}{(z_2)^2} \Bigg \{ \frac{\left[1+\frac{1}{z_1} \delta(\frac{1}{z_1}-\frac{1}{z})\right] \mathrm{Im}\left[\hat{F}_{LT}\left(z_1, z_2\right)\right]}{\frac{1}{z_1}-\frac{1}{z_2}} \nonumber \\
& -\frac{\left[\frac{3}{z_1}-\frac{1}{z_2}+\frac{1}{z_1}(\frac{1}{z_1}-\frac{1}{z_2}) \delta(\frac{1}{z_1}-\frac{1}{z})\right] \mathrm{Im}\left[\hat{G}_{LT}\left(z_1, z_2\right)\right]}{(\frac{1}{z_1}-\frac{1}{z_2})^2} \Bigg \},
\label{ww3a}
\end{align}

\begin{align}
G_{1LT}^{(1)}(z)= &  \frac{m_q}{M}  \int_z^1 d z_1 \frac{H_{1LT}(z_1 )}{z_1}  + \int_z^1 \frac{d z_1}{z_1} \int_{z_1}^{\infty} \frac{d z_2}{(z_2)^2} \Bigg \{ \frac{ \mathrm{Im}\left[\hat{F}_{LT}(z_1, z_2)\right]}{\frac{1}{z_1}-\frac{1}{z_2}} -\frac{(\frac{3}{z_1}-\frac{1}{z_2}) \mathrm{Im}\left[\hat{G}_{LT}\left(z_1, z_2\right)\right]}{(\frac{1}{z_1}-\frac{1}{z_2})^2} \Bigg \}.
\label{ww3b}
\end{align}

If we consider the production of a tensor-polarized hadron $h$ in the lepton-nucleon collision, namely $l+N \to h+X$, the twist-3 cross sections are dependent on the chosen frame, which is induced by the arbitrariness in the choice of lightcone vectors for distribution and fragmentation correlators. The LIRs we derive can be used to remove the frame dependence of the twist-3 cross sections for this process, such as twist-3 SSAs and double-spin asymmetries.

\section{Summary}
\label{summary}

The tensor-polarized FFs of a spin-1 hadron ($h$) can be measured in the various hard semi-inclusive processes such as $e^+e^-\rightarrow h X$ and $e p \rightarrow eh X$ (SIDIS), and the former process is accessible at BESIII and Belle II, while the latter is possible at JLab and the Electron-Ion Colliders in the US and China.
Inspired by the ongoing measurement of the tensor-polarized FFs for $\phi$  at BESIII,  we investigate the theoretical relations among the tensor-polarized intrinsic, kinematical and dynamical FFs for a  spin-1 hadron in this work. First, the QCD e.o.m. relations are obtained for the tensor-polarized FFs. Second, we derive the operator identities where the nonlocal quark-quark operators are expressed in terms of quark-gluon-quark operators. Three new Lorentz invariance relations (LIRs) are presented for the tensor-polarized FFs, and they can be used to remove the frame dependence of the twist-3 spin observables in the hard semi-inclusive reactions so that Lorentz invariance properties are satisfied. Finally, we also show that the intrinsic and  kinematical twist-3 FFs are expressed in terms of  the twist-2 FFs and the dynamical twist-3 FFs, and the Wandzura-Wilczek-type relations are obtained by neglecting the dynamical FFs. Since the twist-2 FFs are much  easier to be accessed in experiment than the twist-3 ones, one can give a rough estimate for the twist-3 FFs by such relations.
Our results will be valuable for the future experimental measurements and theoretical studies of tensor-polarized FFs.

\section*{Acknowledgments}
We acknowledge useful discussions with Shunzo Kumano, Bernard Pire, Ji Xu and Ya-Teng Zhang. Qin-Tao Song was supported by the National Natural Science Foundation
of China under Grant Number 12005191.



\end{document}